\documentclass[final]{elsart}

\usepackage{natbib}

\def\al{\alpha} \def\be{\beta} \def\ga{\gamma}

\def\la{\lambda}

  \def\mn{{\mu\nu}}

 \def\frac#1#2{{\textstyle{{#1}\over
{#2}}}} 
\def\lsim{\mathrel{\rlap{\lower4pt\hbox{\hskip1pt$\sim$}}
\raise1pt\hbox{$<$}}}
\def\gsim{\mathrel{\rlap{\lower4pt\hbox{\hskip1pt$\sim$}}
\raise1pt\hbox{$>$}}} \def\sqr#1#2{{\vcenter{\vbox{\hrule height.#2pt
\hbox{\vrule width.#2pt height#1pt \kern#1pt \vrule width.#2pt} \hrule
height.#2pt}}}}

 \def\beq{\begin{equation}}
\def\eeq{\end{equation}}
\def\beqa{\begin{eqnarray}} \def\eeqa{\end{eqnarray}}

\def\laq{\raise 0.4 ex \hbox{$<$}\kern -0.8 em\lower 0.62
ex\hbox{$\sim$}} \def\gaq{\raise 0.4 ex \hbox{$>$}\kern -0.7 em\lower
0.62 ex\hbox{$\sim$}}

\begin{document}

\begin{frontmatter}

\title{Current tests of alternative gravity theories: the Modified Newtonian Dynamics case}

\author{Orfeu Bertolami}\ead{\newline orfeu@cosmos.ist.utl.pt}\ead[url]{http://alfa.ist.utl.pt/$\sim$orfeu/homeorfeu.html}, \author{Jorge P\'aramos}\ead{\newline x\underline{ }jorge@fisica.ist.utl.pt}

\address{Instituto Superior T\'ecnico, Departamento de F\'\i sica \\
Av. Rovisco Pais 1, Lisbon, 1049-001, Portugal}

\begin{abstract} We address the possibility of taking advantage of
high accuracy gravitational space experiments in the Solar System and
complementary cosmological tests to distinguish between the usual
general relativistic theory from the alternative modified Newtonian
dynamics paradigm.

\end{abstract}


\end{frontmatter}

\section{Introduction}
\label{sec:intro}

The current observational status indicates that gravitational physics
is in a considerable agreement with Einstein's theory of General
Relativity (GR) \citep{Will05,BPT06}; however, there are some theoretical and experimental issues that shed doubt on its status of the ultimate
description of gravity.

Most theoretical difficulties are related to the strong gravitational
field regime, together with the ensuing spacetime singularities. It is believed that a possible way of overcoming these troublesome questions lies in the
quantization of gravity. Unfortunately, the success of modern gauge
field theories in describing the electromagnetic, weak, and strong
interactions does not yet extend to gravity and, in fact, the
fundamental pillars of physics, Quantum Mechanics and General
Relativity, are incompatible with each other. Furthermore, fundamental
theories that attempt to include gravity lead to new long-range
forces. Even at a purely classical level, Einstein's theory does not
provide the most general way to establish the spacetime metric, and
alternative metric theories that have been put forward allow for
violations of the Equivalence Principle, modification of large-scale
gravitational phenomena and variation of the fundamental
``constants'', which are fairly constrained phenomenologically. Thus, the derived predictions serve well to motivate new
searches for very small deviations from General Relativity; these
quests should clearly include further gravitational experiments in
space, including laser astrometric measurements \citep{ESTEC_lator,TexasStanford_lator,solvang_lator04,Lator01},
high-resolution lunar laser ranging (LLR) \citep{Murphy_etal_2002} and
long range tracking of spacecraft using the formation flight concept,
as proposed \citep{Pioneer} to test the Pioneer
anomaly \citep{Anderson02}; a broader review of fundamental physics
experiments in space can be found elsewhere \citep{bertolami,Matos04}.

From an observational standpoint, recent cosmological observations
indicate that, if GR is indeed correct, then one must also assume that
most of the energy content of the Universe lies in some unknown forms
of dark matter and dark energy that may permeate much, if not all
spacetime.  Indeed, recent Cosmic Microwave Background Radiation
(CMBR) WMAP three year data\citep{WMAP3} tells us that the Universe is
well described by a flat Robertson-Walker metric, with an energy
density of the Universe fairly close to its critical value, $\rho_c
\equiv 3H_0^2/8 \pi G \simeq 10^{-29} g/cm^3$, where $H_0 \simeq
71~km~s^{-1}Mpc^{-1}$ is the Hubble expansion parameter at present.
Furthermore, CMBR, Supernova and large scale structure data are
consistent with each other if, in the cosmic budget of energy, dark
energy corresponds to about $73\%$ of the critical density, while dark
matter to about $23\%$ and baryonic matter  to only about $4\%$.

As is widely known, dark matter was firstly suggested in 1933
by Zwicky, from a study of the observed motion of the peripheral
galaxies of the Coma cluster; this indicated a discrepancy between the
total mass, as inferred from its brightness or total number of
galaxies \citep{Trimble}, which appeared to indicate that the total
amount of mass in the cluster is about 400 times more mass than
expected. This led to the postulation of a yet although unaccounted,
non-luminous matter, which became known as ``dark matter''. This
hypothesis was also supported by the differential rotation of our
galaxy, firstly discussed by Oort in 1927, the flatness of
galactic rotation curves, large scale data and, more recently, by the observation of the so-called ``bullet'' cluster \citep{cluster2}.

Returning to current times, one finds that the amount of
new data and related issues coincides with recent progress in
high-precision measurement technologies for physics experiments in
space, ranging from spacecraft navigation to time transfer, clock
synchronization, weight and length standards. Hence, the concerned
physicist is no longer restrained to mere speculation, but has a
chance of probing crucial matters such as the nature of dark energy
and dark matter, effects of intermediate range forces and, ultimately,
the fundamental nature of gravity. Naturally, the evolving techniques
should allow for an ever increasing precision in measurements, with
associated refinement of viable models, and exclusion of unrealistic
ones. Eventually, it is expected that an ultimate ``theory of
everything'', engulfing both Quantum Mechanics and General Relativity
will finally shed some light over the cosmological issues involving
the origin and destiny of the Universe.  In what follows, we discuss
the current status of one of the most widely discussed alternatives to
GR, namely the Modified Newtonian Dynamics and its underlying
fundamental theory, the so-called Tensorial-Vector-Scalar theory. For the discussion of the implications of scalar tensor theories and modifications designed to mimic dark energy the reader is directed to Refs. \citep{BPT06, bertolami}

\section{Modified Newtonian Dynamics and Tensorial-Vector-Scalar
theory}
\label{MOND and TeVeS}

\subsection{Modified Newtonian Dynamics}
\label{MOND}

Although the current paradigm of GR endowed with dark
matter and dark energy components is highly successful in
describing observations on larger scales, and seeding very enticing theoretical
ideas, it is important to study competing alternatives, in particular those which attempt to account for observations without the dark matter component. For instance, a possibility involves the putative running off the gravitational coupling \citep{bertolami93,bertolami96}. Another, well discussed alternative, goes by the name of Modified Newtonian Dynamics (MOND), and is based in a modification of Poisson's equation for the gravitational potential \citep{Milgrom}. Hence, instead of the traditional form, MOND postulates the following equation:

\beq \nabla \cdot \left( \mu \left( {\nabla \phi \over a_0 } \right)
\nabla \phi \right) = 4 \pi G \rho~~, \eeq

\noindent where $\rho$ is the density of barionic matter and $a_0
\approx 10^{-10}~m~s^{-2}$ is the scale at which Poisson's equation
fails; specifically, one regains the traditional form for
accelerations $ a =-  \nabla \phi \gg a_0$, through the Milgrom function
$\mu$, which values $\mu(x) \approx x$ for $x \ll 1 $, and $\mu(x)
\approx 1 $ for $x \gg 1$.

This phenomenological prescription provides with an alternative
solution for the flattening of the rotational curves of
galaxies. Indeed, far away from the central region, the gravitational
potential should be small enough so that MONDian behaviour takes over,
and a simple algebraic computation shows that, instead of the inverse
square law for gravity, one obtains the desired $r^{-1}$ behaviour,
yielding $v_c(r) $ constant. Furthermore, a direct consequence of this
scenario is that the luminosity scales with $v_C^4$, which is in
agreement with the empirical Tully-Fisher law \citep{Milgrom,TFlaw}.

Given that it was designed to solve these and other issues
specifically, MOND is a purely phenomenological model, thus lacking a
suitable support in the form of a fundamental theory, covariantly
formulated in terms of the variation of a specific action
functional. To avoid this caveat, several attempts to generalize MOND
were undertaken, which have so far settled in what is known as
Tensorial-Vector-Scalar theories.

\subsection{Tensorial-Vector-Scalar theory}
\label{TeVeS}

The proposed Tensorial-Vector-Scalar (TeVeS) \citep{Bekenstein} theory 
assumes, asides
from the metric $g_{\al\be}$ and matter fields $\varphi_i$ (with the
associated energy-momentum tenor $T_{\al\be}$, the presence of a
vector field $U^\al$ and two scalar fields, $\sigma$ and $\phi$. The
vector field is everywhere timelike, with the normalization condition
$U_\al U^\al = -1$; furthermore, the vector field couples with normal
matter through a physical metric $\tilde{g}_{\al\be} = e^{-2\phi}
g_{\al\be} - 2 U_\al U_\be ~sinh(2 \phi)$. The action of the theory is
written as $S = S_G + S_V + S_S + S_M$, with

\beqa S_G & = & \int g^\mn R_\mn (-g)^{1/2} ~d^4 x~~, \\ \nonumber S_V
& = & - {K \over 32 \pi G } \int \left[ g^{\al\be} g^\mn U_{[\al,\mu]}
U_{[\be, \nu]} - {2 \la \over K} (g^\mn U_\mu U_\nu + 1 ) \right]
(-g)^{1/2}~ d^4 x ~~, \\ \nonumber S_S & = & - {1 \over 2 } \int
\left[ \sigma^2 h^{\al\be} \phi_{,\al} \phi_{,\be} + {G \over 2 l^2 }
\sigma^4 F(kG\sigma^2) \right] (-g)^{1/2} d^4 x ~~, \\ \nonumber S_M &
= & \int \mathcal{L} (\varphi_i) (-\tilde{g})^{1/2} ~d^4 x~~, \eeqa

\noindent and $K$, $k$ and $l$ are constants specific of the theory,
$\la$ is a Lagrange multiplier that sets the timelike normalization
condition for the vector field, $F$ is a free function and $h^{\al\be}
= g^{\al \be} - U^\al U^\be$. Notice that there is no kinetic term for
the scalar field $\sigma$, whose equation of motion provides an
algebraic relation between its value, the free function $F$ and the dynamic
scalar field $\phi$.

There are a number of criticisms concerning the action of the theory:
firstly, the presence of a non-dynamic $\sigma$ is somewhat {\it ad hoc},
although it is argued that the scalar action resembles that of a
self-interacting complex scalar field in a particular limit
\citep{Bekenstein}. This suggests that $\sigma$ is not a fundamental
field, but only an artificial tweak of the theory in order to enforce
some specific behaviour. The choice of physical metric $\tilde{g}^{\al\be}$ is also somewhat odd, since it is assumed
that the vector field $U^{\al}$ couples directly with matter. Furthermore, there is no coupling between the scalar fields and matter. Also, asides from the
dynamic and physical metrics, a third tensor $h^{\al\be}$ also
appears, which is somewhat similar to the physical metric, but appears
only in the scalar action; this reveals that the vector field couples
to the non-dynamic scalar field only, another arbitrary feature. Finally, the free function $F$ has no rationale apart from the fact that it is specified
in order to obtain the desired features. In the overall, the action
for TeVeS appears as an odd concatenation of different tensorial,
vector and scalar quantities, with no apparent internal consistency, asides
from the fact that it provides MOND with some more underlying supporting
theory.

The full demonstration of the connection between TeVeS and MOND is outside 
the scope of this text, and can be found elsewhere \citep{Bekenstein}. It can be 
shown that 
MOND behaviour arises if one assumes a spherically symmetric, quasistatic 
case, that is, one endowed with a weak potential and where only slow motion 
occurs. For this identification to be complete, Milgrom's function $\mu$ is obtained 
from a specific choice for the free function $F$ . This yields a rather complex 
form for the latter quantity, having an asymptote when its argument goes 
to unity, and diverging as it approaches infinity. The function $F (x)$ also lacks 
definition when its argument varies between $ 1 \leq x \leq 2 $, thus dividing 
it into two different branches; the branch $x < 1$ is relevant in quasi--stationary 
systems, while $x > 2$ concerns cosmological issues. Given that $ F $ appears 
as a potential function for the non-dynamic field $\sigma$, this increases the difficulty associated with the latter, as previously discussed. 

\section{Testing MOND and TeVeS}
\label{testing}

\subsection{Cosmology}
\label{cosmology}

In what follows, we briefly describe the current status of observations, and 
its implications in testing if MOND and TeVeS can be a viable alternative 
to the standard paradigm of GR and dark matter. Before dwelling into its 
natural testing grounds, we quickly address some cosmological issues, namely 
the current accelerated expansion of the Universe and the initial structure 
formation. These are implemented in the context of TeVeS by assuming that 
the vector field has only a temporal component (due to isotropy) and resorting 
to the $ x > 2 $ branch of the free function $ F (x) $. Also, homogeneity requires 
that 
the scalar field depends solely of the cosmic time $t$ \citep{Bekenstein, Skordis}. 

Regarding the accelerated cosmic expansion, it is found that a cosmological 
constant term can be included in TeVeS by adding an appropriate integration 
constant to the free function $ F $; this has no effect in an astrophysical context, but yields the required repulsive force that counteracts the expected deceleration of the rate of expansion \citep{Bekenstein}. However, its inclusion is purely phenomenological, and offers no explanation as to the value of this quantity, nor it provides any clue concerning the much debated cosmological constant problem.
 
The issue of initial structure formation is more evolved; indeed, standard 
cosmology assumes that structure formation occurred due to the condensation 
of initial fluctuations, enhanced by the presence of dark matter \citep{Trimble}. In 
the case 
of TeVeS, which does not assume this component to be present, the situation is 
still under study, but 
proponents of the theory usually rely on TeVeS also manifesting MOND behaviour 
in the early Universe 
scenario \citep{Skordis}. More specifically, instead of density fluctuations of the 
dark matter 
component, it is argued that fluctuations of the scalar field $\phi$ also give rise to 
baryonic structure 
formation. However, these results have been recently challenged by a numerical 
study \citep{Pointecouteau}. 
Concerning other evolutionary stages of the Universe, namely inflation, the 
radiation and matter eras, it is found that the scalar field does not interfere 
significantly with standard 
cosmology, as long as its magnitude does not 
change abruptly between the end of one era and the onset of the next, and is 
globally between $ 0.0007 < \phi \ll 1 $ \citep{Bekenstein}.

\subsection{Galactic Clusters}
\label{clusters}

Recent observations of galactic clusters have been interpreted as a strong 
indication of the validity of the GR with dark matter paradigm. This assessment is based on the gravitational profile reconstructed from gravitational lensing 
data, which clearly shows two strong peaks, coinciding with the centers of the 
colliding galaxies \citep{cluster1,cluster2}. Indeed, most of the ``visible'' matter in a galaxy is in the form of plasma, with stars and dust accounting for only about $20 \%$ of the total; colliding plasma distributions tend to 
merge, yielding a gravitational profile with a single peak, while ``normal'' matter retains its integrity, so that its gravitational profile should show the 
individual peaks of the colliding objects. Given 
that the reconstituted profile for the two galaxy clusters is in accordance with 
the latter, one concludes that a dark matter component dominates each galaxy's mass
distribution. Since known matter amounts only for a small proportion, dark matter is what accounts for the gravitational profile. 
There are, however, some tenacious efforts towards an alternative interpretation of this result, trying to  implement the MOND hypothesis in this 
context. This resorts to an already known feature of MOND (and, fundamentally, 
from TeVeS): that this theory does not completely rule out the need for dark matter, but still enables a reduction of its proportion by a factor of five (with the remainder possibly being made of neutrinos or some unaccounted ``normal'' 
matter) \citep{Sanders}. Hence, it is suggested that MOND can also account for the twin peaks gravitational profile, by assuming the presence of neutrinos with a mass of approximately $2~ eV$ \citep{Angus1,Angus2}. Two criticisms can be put forward: firstly, the assumed neutrino mass is very close to the known allowed upper bound of $ 0.07~eV < m_\nu < 2.2 ~eV$, so its validity must be reassessed as tighter results become available. Secondly, and possibly more crucially, the reconstruction of the observed gravitational profile based on MOND resorts to various linear combinations of ``MONDian'' potentials $\phi_i$ . This is somewhat 
troublesome, given that the inherent non-linearities of the underlying TeVeS 
theory may indicate that, even if each of the galaxies allows for a MONDian 
behaviour away from its central region, one cannot treat their collision as a 
simple superposition of individual potentials. Moreover, the chosen potentials 
are not fully accounted for, with no clear justification given for the values 
of the fitting parameters or the algebraic form of the potentials themselves. 
Also, uncertainties on the mass distribution, gravitational lensing statistics 
and spatial symmetry of galaxy mergers do not allow for a clear relation 
between the fitting parameter giving the strength of the scalar field $\phi$, and the 
typical value $ k \sim 0.03 $ \citep{Bekenstein}, which appears in the free function $ 
F $ of TeVeS and sets the order of magnitude of this scalar field in a spherical symmetric case, 
through 

\beq \phi(r) = \phi_c - {kGM \over 4 \pi r} ~~, \eeq

\noindent where $\phi_c$ is the cosmological value of $\phi$, to which $\phi(r)$ 
goes asymptotically, as 
the metric goes to its Minkowski form. Furthermore, a thorough analysis of galaxy 
merging in the context of MOND should eventually resort to TeVeS as a more fundamental approach to tackle the complexity of the spatial superposition of two matter distributions. A mere fit of parameters based on plausible, but somewhat arbitrary potentials, offers only a superficial solution .

\subsection{Solar System}
\label{solar system}

It is our opinion that implications of MOND in the Solar System should be better examined. Most probably, this lack of interest is due to the preconception that the condition $ \nabla \phi \ll a_0 $ does not occur within 
the vicinity of the Sun (as can be checked by direct computation), and, as 
a result, MOND-like behaviour should be nonexistent or highly attenuated. 
However, there are two avenues of research that may be pursued: the first 
concerns the study of regions where the above condition may be realistically found, 
that is, by assuming not only a simple central body problem, but also take 
the remaining planets and objects into account. Specifically, one can deal with 
a two-body problem, and analyze the region close to the equilibrium point 
$\nabla \phi = 0 $; a partial numerical study of this issue shows that there is a 
MONDian ``bubble'' of non-negligible size at such location, which could provide interesting clues as to the 
validity of the MOND paradigm \citep{Magueijo}. Although direct effects are below current measurement capabilities, it is found that an object located within this bubble would experience an 
anomalous acceleration which, in principle, could be measured in the near future. 

In parallel, one can assume a simpler central body problem and, while assuming 
that no MONDian 
regime occurs in its vicinity, perturbative effects 
should arise in the standard Scharzschild metric. This is clearly suitable for an 
analysis within the framework of the PPN formalism \citep{Will}. In 
short, a simplified version 
of this formalism states that, in a isotropic reference frame, the expansion of the 
metric to second order in 
the gravitational potential reads 

\beq g_{00} = -1 + {2GM \over r} - \be \left( {2 GM \over r } \right)^2~~~,~~~g_{ii} = 
1 + \ga {2 GM \over 
r}~~. \eeq

\noindent The $PPN$ parameters $\be$ and $\ga$ are related to fundamental 
physical properties of the theory under scrutiny: the latter quantifies the spatial curvature produced by unit 
mass, while the former indicates the amount of non-linearity in the superposition law for gravity. These can be used to compare and distinguish GR from competing theories of gravitation, with the former 
displaying the reference values $\be = \ga = 1$.

It was first thought that such a procedure would yield results which made MOND indistinguishable from GR, that is, that the PPN parameters $\be$ and $\ga$ do not shift from unity \citep{Bekenstein}. However, this assumed that the vector field $U^\al$ had only a non-vanishing temporal component which, by virtue of its normalization condition, tracks the metric element $g^{00}$ and, after some algebraic work, 
leads only to a redefinition of Newton's gravitational constant $G$; this prescription for the vector field is allowed by its equation of motion, and argued as a plausible choice for its form. 
However, a later work \citep{Giannios} showed that one can also 
obtain another solution for the vector equation of motion, where $U^\al$ has also 
a non-vanishing radial coordinate, related to the temporal one through the 
normalization condition. Contrary to the previous null deviations from the GR values for the PPN 
parameters, this yields a case which differs from standard GR\footnote{We have recently and independently obtained a very similar result, unaware that it had been derived earlier.}: 

\beq \be = 1 + {k \over 8 \pi} + {K \over 4} + \phi_c \left(3 + {k \over \pi K} \pm \sqrt
{{2k \over \pi K } +5} \right)~~~~, ~~~~\ga=1~~. \eeq

It should be pointed out that, in our opinion, the result $\be \neq 1$, 
asides from its clear theoretical interest, is more physically motivated that 
the previous $\be = \ga = 1 $ case. This is so because the latter is based on the {\it 
Ansatz} $U^\al = 
(U^0,0,0,0)$, while the former takes $U^\al = (U^0,U^r,0,0)$: while both are 
consistent with the vector field
equation of motion, the purely temporal case seems somewhat artificial. Indeed, 
although the velocity--field of known matter admits only a temporal component, the scalar field $\phi$, 
which pervades all space, has a velocity--field with both temporal and radial components. Hence, a 
sensible assumption seems to be that the vector field also has temporal and radial components. Moreover, if indeed one has $\be \neq 1$, the superposition principle assumed in \cite{Angus1,Angus2} for the summation of potentials is not exact, which further casts doubts concerning the ability of MOND to account for the gravitational profiles of galactic clusters.

Recall that the current experimental data shows an impressive agreement with General
Relativity, with the most stringent bounds

\beq \beta - 1 = (1.2 \pm 1.1) \times 10^{-4}~~~~,~~~~ \gamma - 1 = (2.1 \pm 2.5) 
\times 10^{-5}~~, \eeq

\noindent arising respectively from the Cassini's 2003 radiometric experiment 
\citep{Bertotti} and from 
limits on the Strong Equivalence violation parameter, $\eta \equiv 4\beta-\gamma-3
$, that
are found to be $\eta=(4.4\pm4.5)\times 10^{-4}$, as inferred from LLR
measurements \citep{Williams_Turyshev_Boggs_2004}.

Hence, the assumed value of $k \sim 0.03 $ is immediately excluded, which poses 
an issue concerning the choice of the free function $F$, since this quantity specifies the order of 
magnitude of that parameter of TeVeS. Also, one obtains a weak bound on $K$, which must be smaller than 
$10^{-3}$; unfortunately, there are no other tight constraints on this quantity to be measured against with, since 
the latter is related to the strength of the vector action in the full TeVeS action, and 
the vector field $U^\al$ does not play a dynamical role in cosmology or galactic 
structure. In any case, it is clear that more stringent bounds of $\be$ may render relevant constraints for TeVeS.

\section{Discussion and Conclusions}
\label{conclusions}

In this study we have discussed the current status of the Modified Newtonian Dynamics alternative to General Relativity, based on the Tensorial-Vector-Scalar theory; we address both its theoretical foundations, as well as recent observations that question its applicability, attempting to cast a circumstanced criticism of both. We conclude that, although MOND still remains a logical possibility, it is at the verge of being excluded from the observations, unless one is willing to exploit some of its theoretical oddities, most of which unaccounted for. Hence, although it may still be too early for its utter dismissal as a viable alternative to the paradigm of GR, we believe that the case for MOND has grown increasingly weaker, favouring the more canonical approach that assumes the presence of a dark matter component in the Universe. Hopefully, the future will hold further experimental results and theoretical advances in both competing theories, leading to a more profound understanding of gravity.


\begin{thebibliography}{30}
\expandafter\ifx\csname natexlab\endcsname\relax\def\natexlab#1{#1}\fi
\expandafter\ifx\csname url\endcsname\relax
  \def\url#1{\texttt{#1}}\fi
\expandafter\ifx\csname urlprefix\endcsname\relax\def\urlprefix{URL }\fi

\bibitem[{Anderson et~al.(2002)Anderson, Laing, Lau, Liu, Nieto, and
  Turyshev}]{Anderson02}
Anderson, J.~D., Laing, P.~A., Lau, E.~L., Liu, A.~S., Nieto, M.~M., Turyshev,
  S.~G., 2002. ``{S}tudy of the anomalous acceleration of {P}ioneer 10 and
  11''. Phys. Rev.~(D 65), 082004.

\bibitem[{Angus et~al.(2006{\natexlab{a}})Angus, Famaey, and Zhao}]{Angus1}
Angus, G.~W., Famaey, B., Zhao, H., 2006{\natexlab{a}}. ``{C}an {MOND} take a
  bullet? {A}nalytical comparisons of three versions of {MOND} beyond spherical
  symmetry''. Mon.Not.Roy.Astron.Soc.~(371), 138.

\bibitem[{Angus et~al.(2006{\natexlab{b}})Angus, Famaey, Zhao, and
  Shan}]{Angus2}
Angus, G.~W., Famaey, B., Zhao, H., Shan, H., 2006{\natexlab{b}}. ``{O}n the
  {L}aw of {G}ravity, the {M}ass of {N}eutrinos and the {P}roof of {D}ark
  {M}atter'', astro-ph/0609125, submitted to Astrophys. J. Lett.

\bibitem[{Bekenstein and Magueijo(2006)}]{Magueijo}
Bekenstein, J., Magueijo, J., 2006. ``{M}ond habitats within the solar
  system''. Phys. Rev.~(D 73), 103513.

\bibitem[{Bekenstein(2004)}]{Bekenstein}
Bekenstein, J.~D., 2004. ``{R}elativistic gravitation theory for the {MOND}
  paradigm''. Phys. Rev.~(D 70), 083509.

\bibitem[{Bertolami(2006)}]{bertolami}
Bertolami, O., 2006. ``{D}ark {E}nergy, {D}ark {M}atter and {G}ravity''.
  International Workshop: From Quantum to Cosmos: Fundamental Physics Research
  in Space, Washington, astro-ph/0608276, submitted to Int. J. Mod. Phys. D.

\bibitem[{Bertolami et~al.(2006{\natexlab{a}})Bertolami, de~Matos,
  Grenouilleau, Minster, and Volonte}]{Matos04}
Bertolami, O., de~Matos, C.~J., Grenouilleau, J.~C., Minster, O., Volonte, S.,
  2006{\natexlab{a}}. ``{P}erspectives in fundamental physics in space''. Acta
  Astronautica~(59), 490.

\bibitem[{Bertolami and Garc\'{i}a{-}Bellido(1996)}]{bertolami96}
Bertolami, O., Garc\'{i}a{-}Bellido, J., 1996. ``{A}strophysical and
  cosmological constraints on a scale dependent gravitational coupling''. Int.
  J. Mod. Phys.~(D 5), 363.

\bibitem[{Bertolami et~al.(1993)Bertolami, Mour{\~a}o, and
  Perez{-}Mercader}]{bertolami93}
Bertolami, O., Mour{\~a}o, J.~M., Perez{-}Mercader, J., 1993. ``{Q}uantum
  gravity and the large scale structure of the universe''. Phys. Lett.~(B 311),
  27.

\bibitem[{Bertolami et~al.(2006{\natexlab{b}})Bertolami, P\'aramos, and
  Turyshev}]{BPT06}
Bertolami, O., P\'aramos, J., Turyshev, S.~G., 2006{\natexlab{b}}. ``{G}eneral
  {T}heory of {R}elativity: {W}ill it survive the next decade?''. 359th
  WE-Heraeus Seminar: ``Lasers, Clock, and Drag-Free: Technologies for Future
  Exploration in Space and Gravity Tests'', University of Bremen, ZARM.

\bibitem[{Bertotti et~al.(2003)Bertotti, Iess, and Tortora}]{Bertotti}
Bertotti, B., Iess, L., Tortora, P., 2003. ``{A} test of {G}eneral {R}elativity
  using radio links with the {C}assini spacecraft''. Nature~(425), 374.

\bibitem[{Clowe et~al.(2006)Clowe, Bradac, Gonzalez, Markevitch, Randall,
  Jones, and Zaritsky}]{cluster2}
Clowe, D., Bradac, M., Gonzalez, A.~H., Markevitch, M., Randall, S.~W., Jones,
  C., Zaritsky, D., 2006. ``{A} direct empirical proof of the existence of dark
  matter'', astro-ph/0608407, submitted to Astrophys. J. Lett.

\bibitem[{Gavazzi et~al.(2003)Gavazzi, Fort, Mellier, Pello, and
  Dantel-Fort}]{cluster1}
Gavazzi, R., Fort, B., Mellier, Y., Pello, R., Dantel-Fort, M., 2003. ``{A}
  radial mass profile analysis of the lensing cluster {MS}2137-23''.
  Astron.Astrophys.~(403), 11.

\bibitem[{Giannios(2005)}]{Giannios}
Giannios, D., 2005. ``{S}pherically symmetric, static spacetimes in
  {T}e{V}e{S}''. Phys. Rev.~(D 71), 103511.

\bibitem[{{LATOR Collaboration (Turyshev et al.)}(2005)}]{ESTEC_lator}
{LATOR Collaboration (Turyshev et al.)}, 2005. ``{F}undamental {P}hysics with
  the {L}aser {A}strometric {T}est of {R}elativity''. ESA Spec. Publ.~(588),
  11.

\bibitem[{Milgrom(1983)}]{Milgrom}
Milgrom, M., 1983. ``{A} {M}odification of the {N}ewtonian dynamics as a
  possible alternative to the hidden mass hypothesis''. Astrophys. J.~(270),
  365.

\bibitem[{{Murphy et al.}(2002)}]{Murphy_etal_2002}
{Murphy et al.}, 2002. ``{T}he {A}pache {P}oint {O}bservatory {L}unar
  {L}aser-ranging {O}peration ({APOLLO})''. 12th International Workshop on
  Laser Ranging, Matera, Italy.

\bibitem[{{Pioneer Colaboration (Dittus et al.)}(2005)}]{Pioneer}
{Pioneer Colaboration (Dittus et al.)}, 2005. ``{A} {M}ission to {E}xplore the
  {P}ioneer {A}nomaly''. ESA Spec. Publ.~(588), 3.

\bibitem[{Pointecouteau(2006)}]{Pointecouteau}
Pointecouteau, E., 2006. ``{T}he mass missing problem in clusters: {D}ark
  matter or modified dynamics?'', astro-ph/0607142.

\bibitem[{Sanders(1998)}]{Sanders}
Sanders, R., 1998. ``{R}esolving the virial discrepancy in clusters of galaxies
  with modified newtonian dynamics'', astro-ph/9807023.

\bibitem[{Skordis et~al.(2006)Skordis, Mota, Ferreira, and Boehm}]{Skordis}
Skordis, C., Mota, D.~F., Ferreira, P.~G., Boehm, C., 2006. ``{L}arge scale
  structure in {B}ekenstein's theory of relativistic {MOND}''. Phys. Rev.
  Lett.~(96), 011301.

\bibitem[{{Spergel et al.}(2006)}]{WMAP3}
{Spergel et al.}, 2006. ``{W}ilkinson {M}icrowave {A}nisotropy {P}robe ({WMAP})
  {T}hree {Y}ear {R}esults: Implications for {C}osmology'', astro-ph/0603449,
  submitted to Astrophys. J.

\bibitem[{Trimble(1987)}]{Trimble}
Trimble, V., 1987. ``{E}xistence and nature of dark matter in the {U}niverse''.
  Ann.Rev.Astron.Astrophys.~(25), 425.

\bibitem[{Turyshev et~al.(2004{\natexlab{a}})Turyshev, Shao, and
  K.~L.~Nordtvedt}]{TexasStanford_lator}
Turyshev, S.~G., Shao, M., K.~L.~Nordtvedt, J., 2004{\natexlab{a}}. ``{T}he
  {XXII} {T}exas {S}ymposium on {R}elativistic {A}strophysics''. Stanford
  University, December, pp. SLAC--R--752.
\newline\urlprefix\url{http://www.slac.stanford.edu/econf/C041213/}

\bibitem[{Turyshev et~al.(2004{\natexlab{b}})Turyshev, Shao, and
  Nordtvedt}]{solvang_lator04}
Turyshev, S.~G., Shao, M., Nordtvedt, K.~L., 2004{\natexlab{b}}.
  ``{E}xperimental design for the {LATOR} mission''. Int.J.Mod.Phys.~(D 13),
  2035.

\bibitem[{Turyshev et~al.(2004{\natexlab{c}})Turyshev, Shao, and
  Nordtvedt}]{Lator01}
Turyshev, S.~G., Shao, M., Nordtvedt, K.~L., 2004{\natexlab{c}}. ``{T}he
  {L}aser {A}strometric {T}est {O}f {R}elativity mission''. Class. Quant.
  Grav.~(21), 2773.

\bibitem[{Will(2001)}]{Will}
Will, C.~M., 2001. ``{T}he {C}onfrontation between {G}eneral {R}elativity and
  experiment''. Living Rev. Rel.~(4), 4.

\bibitem[{Will(2005)}]{Will05}
Will, C.~M., 2005. ``{W}as {E}instein {R}ight? {T}esting {R}elativity at the
  {C}entenary''. ``100 Years of Relativity: Spacetime Structure - Einstein and
  Beyond'', World Scientific, Singapore.

\bibitem[{Williams et~al.(2004)Williams, Turyshev, and
  Boggs}]{Williams_Turyshev_Boggs_2004}
Williams, J.~G., Turyshev, S.~G., Boggs, D.~H., 2004. ``{P}rogress in lunar
  laser ranging tests of relativistic gravity''. Phys. Rev. Lett.~(93), 261101.

\bibitem[{Zhao and Famaey(2006)}]{TFlaw}
Zhao, H.~S., Famaey, B., 2006. ``{R}efining {MOND} interpolating function and
  {T}e{V}e{S} {L}agrangian''. Astrophys. J.~(638), L9.

\end{thebibliography}
\end{document}